# Simulation dose enhancement in radiotherapy caused by cisplatin

A.A. Baulin[a,b,1], L.G. Sukhikh[a] and E.S. Sukhikh[a,c]

[a]*National Research Tomsk Polytechnic University, Tomsk, 634050, Russian Federation*
[b]*Gamma Clinic High-Precision Radiology Centre (Gamma Medtechnology Ltd), Obninsk, 249034, Russian Federation*
[c]*Tomsk Regional Oncology Centre, Tomsk, 634050, Russian Federation*

*E-mail: baylin1991@tpu.ru*

ABSTRACT: This research considers potential dose increase in target due to cisplatin (Pt) concentration and radiation type. Cisplatin concentrations from 0.003 to 120 mM were used. Monte-Carlo simulation of Linear accelerator (Elekta Synergy) and X-ray tube (Xstrahl300) was carried out using Geant4 and PClab. As the first step of this research, we performed simulation of energy spectrum from radiotherapy units (spectrum model). The next step was the modeling of linear accelerator head and X-ray tube, and the distribution of dose in the water phantom (PDD model). At the second stage, dose changes were investigated in the presence of cisplatin in the target (CIS model). The simulation results showed that the dose escalation can be caused by photon-capture therapy (PCT). There is a dose enhancement in the volume where cisplatin is accumulated. Then higher is concentration, then higher is the effect. However, the photon energy increase from 60 to 250 kV and increase of depth of target reduces the effect of PCT due the decrease of the photoelectric effect cross-section. Should be noticed, that the orthovoltage X-rays energy, listed in the table with results shows higher dose enhancement, than the megavoltage photon beam generated from linear acceleration sources. In addition, that the dose enhancement factors (DEF) are higher in linac without flattening filter, than in linac with flattening filter.

---

[1] Corresponding author.

**Contents**

**1 Introduction**

Radiation therapy is an important and effective option for treatment of malignant tumors [1]. Actual problem of radiotherapy is increasing effectiveness and reducing side effects of the treatment. Binary technologies of radiation therapy can be used to improve treatment results. One of the most promising technology is contrast-enhanced or "photon-capture" radiotherapy (PCT) [2]. The basic principle of PCT is the generation of a large number of the characteristic X-rays and low-energy Auger-electrons due to interaction between photons and nuclei of heavy elements ($Z \geq 53$). In biological tissue this secondary low-energy radiation ionizes nearby atoms and leads to the occurrence of highly active radical series, which causes the destruction of the macromolecules of DNA and RNA as well as other cell structures.

    The PCT is effective at low X-ray energies where the photoelectric effect dominates (up to about 200 kV) [3]. Accordingly, the energy escalation of photon beams leads to a gradual decrease the effect of the PCT. However, low-energy photons for medium- and deep-seated tumors are limited due to their low penetrating ability. But, these orthovoltage X-rays energy may be used in an intraoperative therapy or to treatments superficial lesions. And for irradiation of deep-seated tumors, the megavoltage photon beams from linear accelerators are widely used, where the Compton effect is maximal and the photoelectric effect is minimal. Although the effect of PCT should be low, different studies show the escalation of energy in the target volume due to the introduction of dose-enhancing agent (DEA) at megavolt photon beams [4,5,6]. Some authors suggest that the observed effect is caused by the wide energy spectrum of photon beams, which including the low-energy kilovolt energy range.

    In this study the chemotherapeutic drug cisplatin ($Cl_2H_6N_2Pt$), which contains platinum atoms, was taken as DEA. Cisplatin, a widely used cytostatic drug, causes cell cycle arrest, inhibition and transcription and, ultimately, apoptosis, i.e. cell death [7]. In the chemoradiation method, the use of cisplatin showed that the drug has not only a chemical effect, but also an obvious radiosensitization effect in the tumor [8, 9].

In this research we simulated the dependence of dose increase on the DEA concentration at 6 and 10 MV medical Linac and orthovoltage X-rays radiation using Monte-Carlo simulation by the means of Geant4.

## 2 Materials and methods

### 2.1 Clinical Dosimetry

The percentage depth dose (PDD) curves were measured on the 6 and 10MV linear accelerator (Elekta Synergy, Elekta Ltd.) and X-ray tube (Xstrahl300) with energy 60, 120, 180 and 250 kV (figure 1).

The PDD of Linear accelerator in a water phantom (Blue Phantom, IBA) using ionization chamber CC13(IBA) having a chamber volume was 0.13 $cm^3$, was measured with 1 mm step. The source-surface distance (SSD) was 100 cm, the radiation field area was 10×10 $cm^2$. And the PDD in a water phantom by 1 mm increments from Xstral300 X-ray tube was measured.

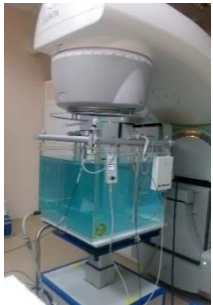
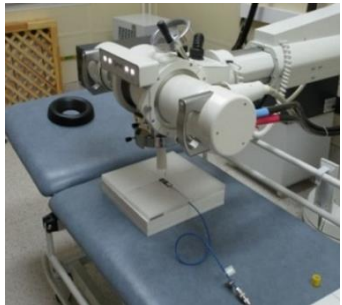

a. Elekta Synergy linac     b. Xstrahl300 X-ray tube
Figure 1. The PDD measurement

Dose measurements in the energies range from 100 kV to 250 kV using an ionization chamber PPC40 (IBA) having a chamber volume was 0.40 $cm^3$, were carried out. The 60 kV energy were measured using the SP34 QA phantom and farmer chamber PTW 23342 having a chamber volume was 0.02 $cm^3$. The experimental measurements were compared with that calculated using Monte-Carlo simulation in water.

### 2.2 Monte-Carlo simulation

Monte-Carlo simulation was carried out using Geant4 version 10 [10] and PClab version 9.9 [11] codes to perform the dose calculation in this research. The first step of this research we performed simulation of energy spectrum from radiotherapy units (spectrum model). The next step is the modeling of head of linear accelerator and X-ray tube and the distribution of dose in the water phantom (PDD model). At the second stage, dose changes were investigated in the presence of cisplatin in the target (CIS model). A total of 1 x $10^8$ histories was run in the model's calculation and the statistical uncertainty of the simulation was kept less than 1 %.

### 2.3 Medical Linac simulation

Figure 2a shows the spectrum model structure considered in this study. Simulated components included: source of electron beams, X-ray target 1mm thick, copper holder below the target 4mm thick and sensitive detector 1nm thick. The source was modeled in a vacuum space. Electron beam was 2 mm in diameter. Accelerated electron beam was bombarded to the tungsten target to produce photon beam. For the nominal 6 MV energy photon beam, the incident electron beam with a mean energy of 6.7 MeV and a Gaussian energy spread of 0.2 MeV were used. The focal spot size was 3.0 mm in the crossline direction. For the 10 MV energy photon beam the corresponding values were 10.4 MeV, 0.3 MeV and 3.0 mm.

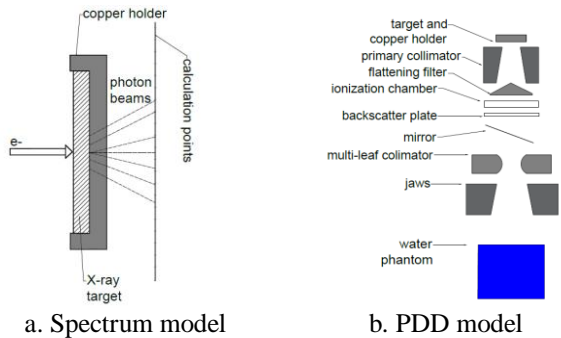

| a. Spectrum model | b. PDD model |

Figure 2. Monte-Carlo simulation of Linac

Based on the manufacturer specifications, we simulated the head of the medical linear accelerator Elekta Synergy located at the Tomsk Regional Oncology Centre. Figure 2b present the head structure of the linear accelerator considered in this study (PDD model). Simulated components included: X-ray target, Primary conical collimator, X-ray beam flattening filter, ionization chamber, thin mylar mirror, Multi-Leaf Collimator (MLC), Asymmetric jaws.

The code of PDD simulation from medical linear accelerator was calculation for filtered and flattening filter free (FFF) systems. The distance between the accelerator source and the water phantom surface (SSD) was 100 cm. Square field 10x10 cm$^2$ were studied. The voxel size of sensitive detector was 0.5x0.5x0.1 cm$^3$.

## 2.4 X-ray tube simulation

In result of interaction between beams of primary particles (electrons) and the tungsten anode of the X-ray tube is generated to bremsstrahlung and characteristic radiation. Energy spectrum of generated X-ray photon beams in detector 1 nm thick was calculated. Figure 3a shows a simulation of the generating x-ray photons (spectrum model). The electron beam is multidirectional point source and diameter is equal to 2.00 mm. The anode is located at an angle of 20 degrees. The model includes a beryllium window 2 mm thick.

At the second stage, the interaction of X-ray photons with water phantoms was simulated to obtain data on depth-dose distributions. Figure 3b shows a simulation of PDD in water phantom (PDD model). The model includes primary filters is half-value layer (HVL), additional filter of various thicknesses, conical applicator and water phantom.

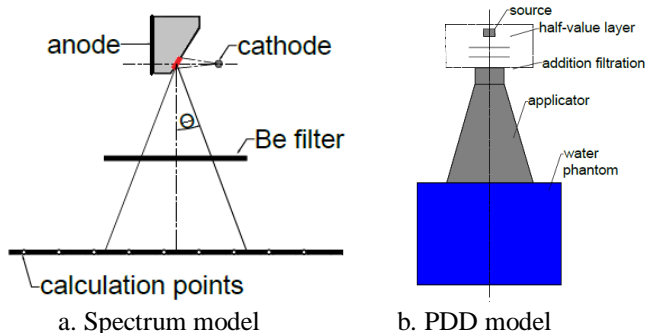

| a. Spectrum model | b. PDD model |

Figure 3. Monte-Carlo simulation of X-ray tube

Percent depth dose with sensitive ring detector was defined 1 cm radius and 1 mm thick. The PDD with a 1 mm step were scored using detector inside a water phantom with dimensions was 41x47,8 см$^2$ at the source to surface distance is equal 50 cm for energy more than 100 kV and 30 cm for energy up to 100 kV. The radiation field with a transverse size of 10x10 cm$^2$ was used.

The graphic image of the models is presented to figure 4 for Xstral300 X-ray tube and figure 5 for Elekta Synergy linear accelerator.

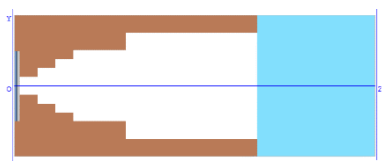 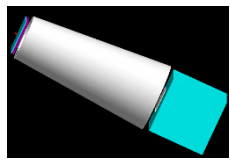 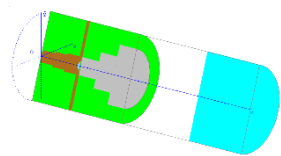 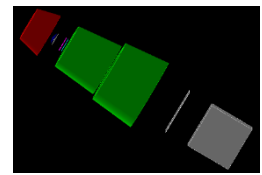

| a. Pclab | b. Geant4 | a. Pclab | b. Geant4 |

Figure 4. The graphic image of Xstral300 simulation

Figure 5 The graphic image of Elekta Synergy simulation

## 2.5 Cisplatin simulation (CIS model)

In the next stage of dose change simulation, we performed with the presence of cisplatin in the target volume. Cisplatin concentrations were selected on the basis of acceptable doses that were considered in previous works by the authors and higher to evaluate dose changes from cisplatin concentration [12,13,14,15,16], while 0,003 mM is the minimum cisplatin concentration during which there is a visible effect in radiobiological experiments after X-ray irradiation. Cisplatin was injected into the target, which was located at a certain depth of water phantom. A 1 mm, 10 mm, 30 mm and 50 mm depths of target location for X-ray tube simulation and 0.5 mm, 50 mm depths of target location for linac simulation were considered. This allowed us to observe not only the dependence on DEA concentration, but also the dependence on the depth of the target location. The calculations of the cisplatin simulation with DEA was carried out while keeping absolutely everyone the parameters of the PDD model concept.

## 3 Results

### 3.1 Spectrum model

The photon energy spectrum as a function of photon energy for linac is shown in figure 6. The energy spectra of incident photons peak were found at 0.511 MV for the maximum nominal energy is 6 and 10 MV.

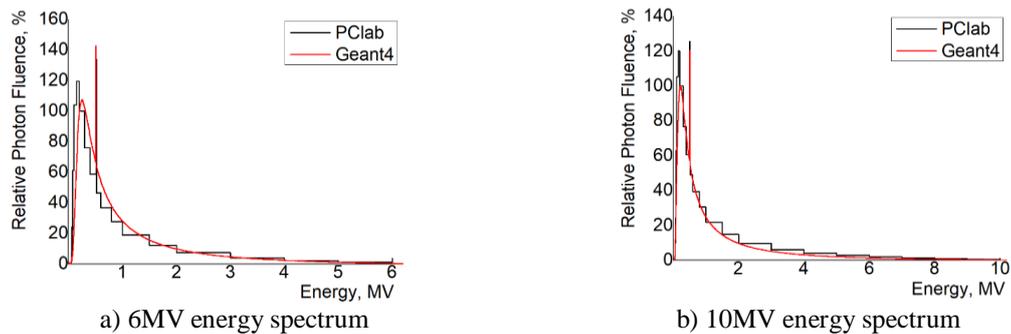

a) 6MV energy spectrum  
b) 10MV energy spectrum  
Figure 6. The photon beams energy spectrum after X-ray target

The photon beam energy spectra in the tube voltages including 60, 120, 180 and 250 kV by the Geant4 and PClab were calculated. The results are presented in figures 7 to 10.

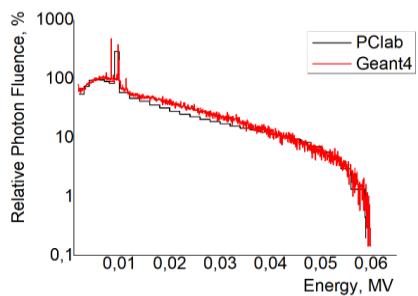 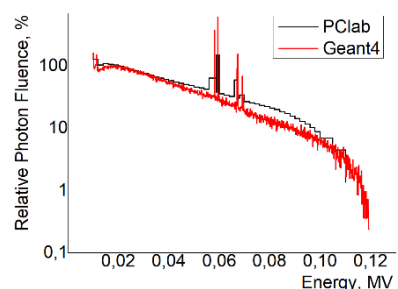

Figure 7. Comparison of X-ray spectra calculation using the Geant4 and PClab for 60 kV tube voltage

Figure 8. Comparison of X-ray spectra calculation using the Geant4 and PClab for 120 kV tube voltage

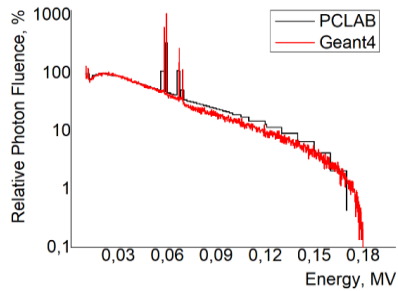
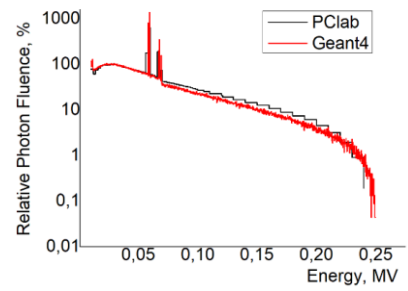

Figure 9. Comparison of X-ray spectra calculation using the Geant4 and PClab for 180 kV tube voltage

Figure 10. Comparison of X-ray spectra calculation using the Geant4 and PClab for 250 kV tube voltage

The results for these energies showed a similar mathematical differences between Geant4 and PClab calculations. The small differences can be associated to the different algorithms and cross-section files used for different systems. Nevertheless, the results confirmed that spectrum model is accurate and can be used for dose distribution calculations in water phantom.

### 3.2 Percentage depth dose model

The PDD obtained by simulations were compared with experimental data obtained with the linear accelerator Elekta Synergy in a Tomsk Regional Oncology Center. The PDD for the 10x10 cm2 irradiation area is shown in figure 11 for 6 MV and in figure 12 for 10 MV. The dose maximum point was determined at 1.5 cm depth for 6 MV and 2.2 cm depth for 10MV. The theoretical value of the maximum depth of the dose is 1.5 cm for 6 MV and 2.3 cm for 10 MV.

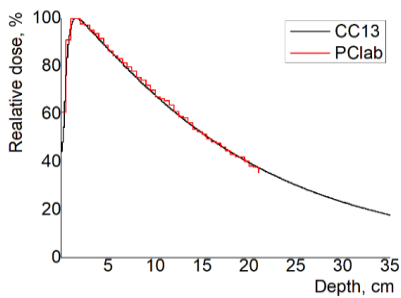
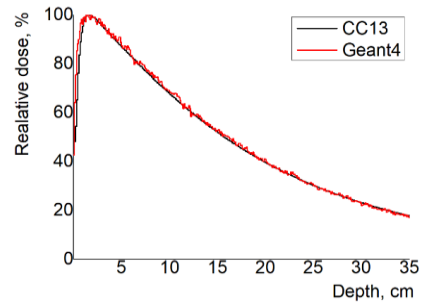

a) PClab calculation and measured of PDD  
b) Geant4 calculation and measured of PDD  
Figure 11. Comparison of PDD calculation with experimentally measured using CC13 for 6 MV linear accelerator

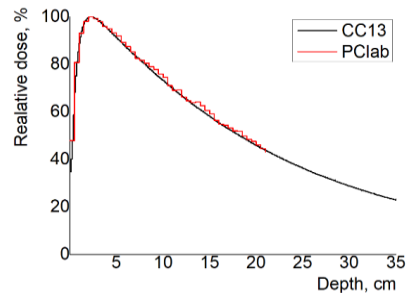
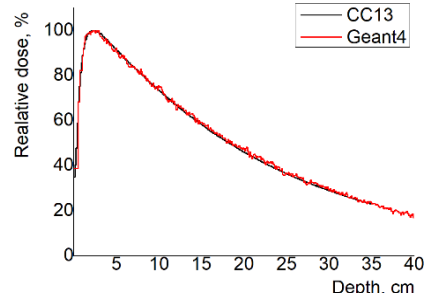

a) PClab calculation and measured of PDD  
b) Geant4 calculation and measured of PDD  
Figure 12. Comparison of PDD calculation with experimentally measured using CC13 for 10 MV linear accelerator

Figure 13 shows the deviation between simulation and measured results. The mean (maximum) deviation was 0.42 % (0.87 %) and 0.47 % (0.80 %) for PClab and Geant4. Figures 14–17 show comparison of PDD obtained from experimental measurements using of ionization chamber and calculated in the Geant4 and PClab. The curves was 60, 120, 180 and 250 kV tube voltage matched very well within 2 % of PDD values at all depths except 1.21cm depth and 120kV where the difference of 2.7 % at PClab and 3.8 % at Geant4 from experimentally measured.

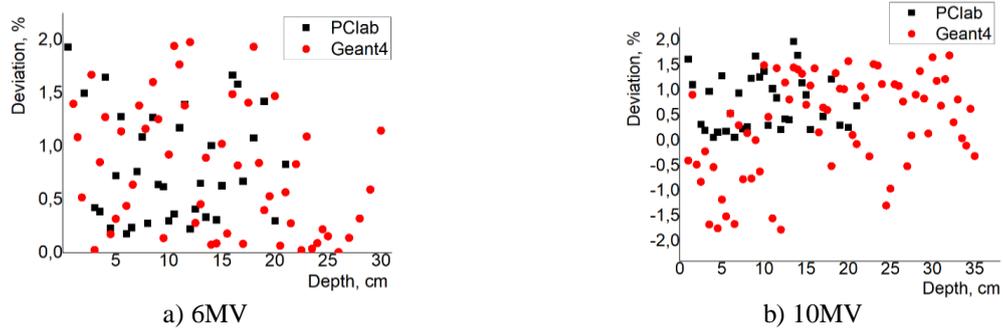

a) 6MV  b) 10MV

Figure 13. The PDD difference simulated by Geant4 and PClab

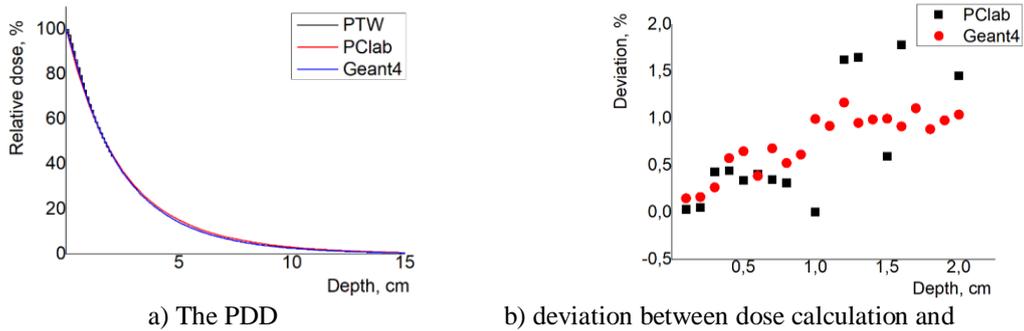

a) The PDD  b) deviation between dose calculation and experimentally measured

Figure 14. Comparison of PDD calculation using the Geant4 and PClab with experimentally measured for 60 kV tube voltage.

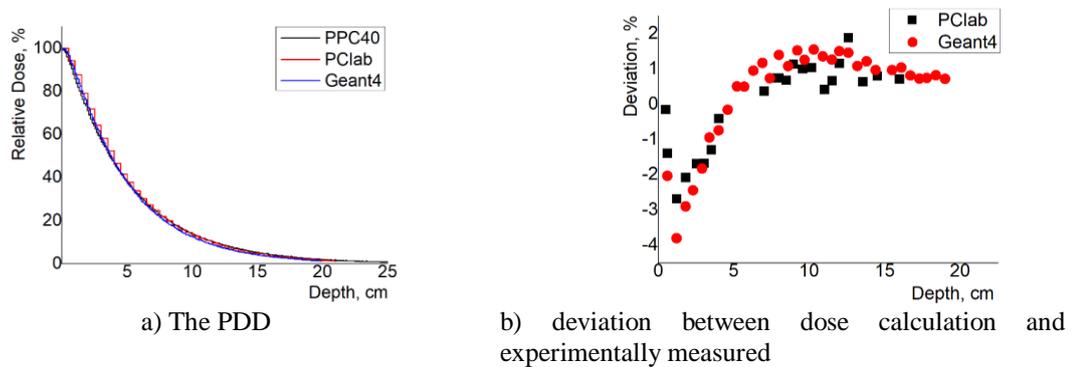

a) The PDD  b) deviation between dose calculation and experimentally measured

Figure 15. Comparison of PDD calculation using the Geant4 and PClab with experimentally measured for 120 kV tube voltage.

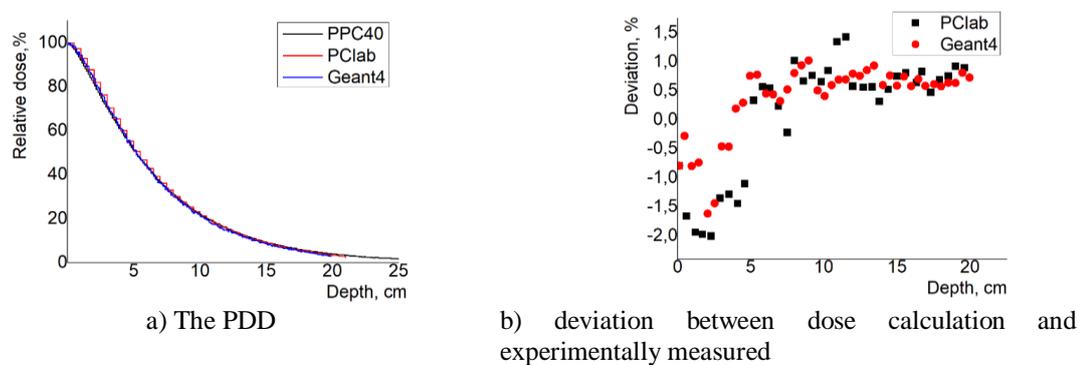

a) The PDD  b) deviation between dose calculation and experimentally measured

Figure 16. Comparison of PDD calculation using the Geant4 and PClab with experimentally measured for 180 kV tube voltage.

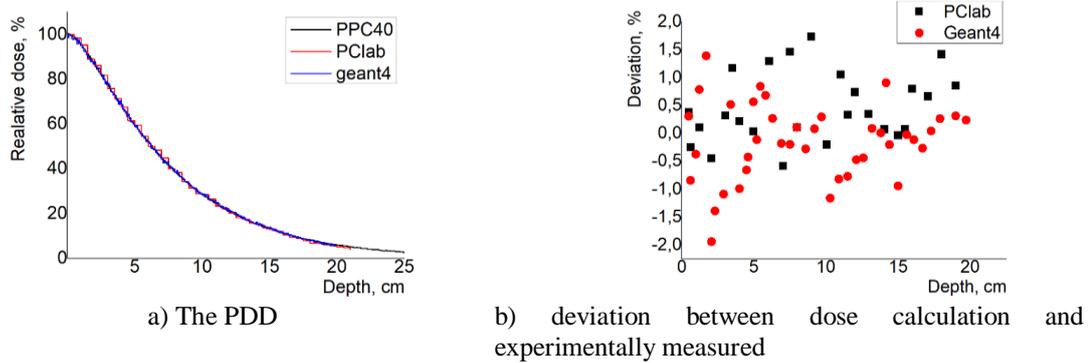

| a) The PDD | b) deviation between dose calculation and experimentally measured |

Figure 17. Comparison of PDD calculation using the Geant4 and PClab with experimentally measured for 250 kV tube voltage.

### 3.3 CIS model

In the next stage the simulation of dose change we performed in the presence of cisplatin in the target volume at Linac and X-ray tube irradiation. Table 1–5 presents the dose-enhancing factor (DEF) for different depth target location and different cisplatin concentration were calculated by Geant4 and PClab. Each of the cases in this table that the no DEA outside the tumor volume was supposed. It's in order to provide a clear relationship between the DEA concentration and beam energy.

The dose enhancement factors (DEF), defined as the ratio of the dose in the tumor volume ($D_a$) with DEA to that dose in the tumor without DEA ($D$):

$$DEF = \frac{D_a}{D}$$

In case of linear accelerator simulation the 5 and 50 mm depths of target location and energy photon beam to 6 MV and 10 MV were considered. This allowed us to observe not only the dependence on agent concentration, but also the dependence on the depth of the target location and energy of photons. In addition, linac without a flattening filter (FFF) were considered. The table 1 shows the dose-enhancing factor for reference points at linacs with flattening filter and flattening filter free (FFF).

As a result, the DEFs calculation for linac does not observe clinically important dose increase. At a concentration of 120 mM, a dose increase of up to 7.7 % and 3.1 % is observed for 6 MV and 10 MV correspondingly. As well as studies of flattening filter free showed the DEF increase equal to 8.6 % and 8.9 % when irradiated with a photon beams of 6 and 10 MV (Figure 18).

Table 1. The DEF values for linac in the energy photon beam equal to 6MV and 10MV. 6MV and 10MV – Linac with flattening filter; 6MVFFF and 10MVFFF - Linac flattening filter free.

| Conc. mM | 6MV | | 10MV | | 6MV FFF | | 10MV FFF | |
|---|---|---|---|---|---|---|---|---|
| 0.5cm | PC | G4 | PC | G4 | PC | G4 | PC | G4 |
| 12 | 1.0062 | 1.0046 | 1.0000 | 1.0000 | 1.0012 | 1.0042 | 1.0022 | 1.0000 |
| 120 | 1.0200 | 1.0768 | 1.0208 | 1.0314 | 1.0327 | 1.0861 | 1.0241 | 1.0557 |
| 5cm | PC | G4 | PC | G4 | PC | G4 | PC | G4 |
| 12 | 1.0000 | 1.0047 | 1.0038 | 1.000 | 1.0013 | 1.0141 | 1.000 | 1.0024 |
| 120 | 1.0659 | 1.0783 | 1.0102 | 1.0205 | 1.0266 | 1.0858 | 1.0216 | 1.0887 |

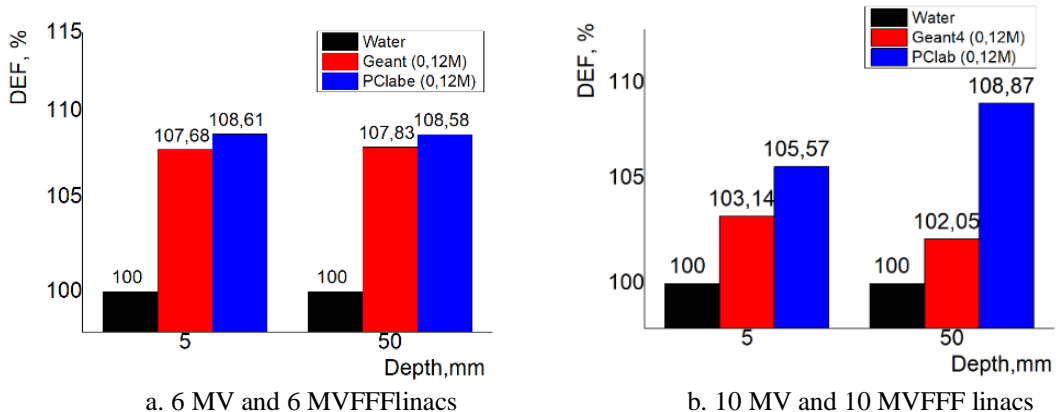

| a. 6 MV and 6 MVFFFlinacs | b. 10 MV and 10 MVFFF linacs |

Figure 18. Comparison of DEF from a linear accelerator at a cisplatin 120 mM concentration

In case of X-ray tube the 1 mm, 10 mm, 30 mm and 50 mm depths of target location and energy photon beam equal to 60, 120, 180 and 250 kV were considered. The tables 2-5 shows the DEF for reference points. From the tables it can be noted, that the DEF increases for higher values in the presence of the cisplatin. There is a dose enhancement in the volume where cisplatin is accumulated. The higher is concentration, the higher is the effect. However, the photon energy increase from 60 to 250 kV reduces the effect of PCT due the decrease of the photoelectric effect cross-section. The highest DEF value is observed when irradiated with 60 kV x-ray.

Table 2. The DEF for 0.1 mm depth target location

| Energy, kV | 60 | | 120 | | 180 | | 250 | |
|---|---|---|---|---|---|---|---|---|
| Conc., mM | PClab | Geant4 | PClab | Geant4 | PClab | Geant4 | PClab | Geant4 |
| 0,003 | 1.0006 | 1,0012 | 1,0000 | 1.0000 | 1.0000 | 1.0000 | 1.0000 | 1.0000 |
| 0,006 | 1.0005 | 1,0008 | 1,0000 | 1.0000 | 1.0000 | 1.0000 | 1.0000 | 1.0000 |
| 0,009 | 1.0003 | 1,0005 | 1.0020 | 1.0024 | 1.0047 | 1.0029 | 1.0029 | 1.0030 |
| 0,012 | 1.0015 | 1,0019 | 1.0025 | 1.0030 | 1.0140 | 1.0084 | 1.0074 | 1.0014 |
| 0,3 | 1.0086 | 1.0138 | 1.0121 | 1.0107 | 1.0166 | 1.0133 | 1.0068 | 1.0052 |
| 0,6 | 1.0186 | 1.0153 | 1.0130 | 1.0146 | 1.0144 | 1.0156 | 1.0044 | 1.0047 |
| 0,9 | 1.0260 | 1.0272 | 1.0233 | 1.0220 | 1.0187 | 1.0213 | 1,0092 | 1.0130 |
| 1,2 | 1.0343 | 1.0293 | 1.0252 | 1.0272 | 1.0229 | 1.0257 | 1.0012 | 1.0013 |
| 3 | 1.0851 | 1.0842 | 1.0771 | 1.0772 | 1.0591 | 1.0587 | 1.0302 | 1.0317 |
| 6 | 1.1704 | 1.1702 | 1.1509 | 1.1510 | 1.1033 | 1.1070 | 1.0638 | 1.0643 |
| 9 | 1.2557 | 1.2554 | 1.2238 | 1.2226 | 1.1714 | 1.1733 | 1.1172 | 1.1124 |
| 12 | 1.3396 | 1.3268 | 1.2860 | 1.2916 | 1.2294 | 1.2278 | 1.1491 | 1,1404 |
| 30 | 1.8210 | 1.8162 | 1.7218 | 1.7394 | 1.5489 | 1.5343 | 1.3337 | 1,3256 |
| 60 | 2.5988 | 2.5800 | 2.4155 | 2.4373 | 2.0902 | 2.0780 | 1.6522 | 1,6665 |
| 90 | 3.3372 | 3.3146 | 3.1095 | 3.1361 | 2.6980 | 2.6644 | 1.9760 | 1,9834 |
| 120 | 4.0348 | 4.0013 | 3.8555 | 3.9135 | 3.1744 | 3.1220 | 2.2702 | 2.2866 |

Table 3. The DEF for 10 mm depth target location

| Energy, kV | 60 | | 120 | | 180 | | 250 | |
|---|---|---|---|---|---|---|---|---|
| Conc., mM | PClab | Geant4 | PClab | Geant4 | PClab | Geant4 | PClab | Geant4 |
| 0,003 | 1.0012 | 1.0001 | 1.0000 | 1.0000 | 1.0000 | 1.0000 | 1.0000 | 1.0000 |
| 0,006 | 1.0000 | 1.0001 | 1.0001 | 1.0000 | 1.0000 | 1.0000 | 1.0000 | 1.0000 |
| 0,009 | 1.0007 | 1.0014 | 1.044 | 1.0000 | 1.0011 | 1.0000 | 1.0000 | 1.0000 |
| 0,012 | 1.0000 | 1.0000 | 1.0875 | 1.0000 | 1.0012 | 1.0009 | 1.0000 | 1.0000 |
| 0,3 | 1.0148 | 1.0099 | 1.0090 | 1.0092 | 1.0111 | 1.0183 | 1.0035 | 1.0035 |
| 0,6 | 1.0176 | 1.0169 | 1.0161 | 1.0150 | 1.0097 | 1.0246 | 1.0084 | 1.0072 |
| 0,9 | 1.0293 | 1.0250 | 1.0270 | 1.0244 | 1.0211 | 1.0314 | 1.0175 | 1.0164 |
| 1,2 | 1.0305 | 1.0336 | 1.0287 | 1.0307 | 1.0144 | 1.0323 | 1.0126 | 1.0138 |
| 3 | 1,0842 | 1.0860 | 1.0738 | 1.0717 | 1.0565 | 1.0610 | 1.0365 | 1.03512 |
| 6 | 1.1739 | 1.1728 | 1.1519 | 1.1493 | 1.1126 | 1.1216 | 1.0557 | 1.0568 |
| 9 | 1.2611 | 1.2588 | 1.2247 | 1.2237 | 1.1753 | 1.1798 | 1.0998 | 1.0986 |
| 12 | 1.3348 | 1.3438 | 1.2869 | 1.2973 | 1.2165 | 1.2372 | 1.1410 | 1.1407 |

| | | | | | | | |
|---|---|---|---|---|---|---|---|
| 30 | 1.8477 | 1.8555 | 1.7324 | 1.7385 | 1.5681 | 1.5684 | 1.3198 | 1.3189 |
| 60 | 2.6661 | 2.6910 | 2.4120 | 2.4321 | 2.1225 | 2.1228 | 1.6669 | 1.6670 |
| 90 | 3.48042 | 3.5167 | 3.1560 | 3.1822 | 2.6846 | 2.6771 | 2.0013 | 2.0014 |
| 120 | 4.2669 | 4.3215 | 3.8518 | 3.8963 | 3.2374 | 3.2360 | 2.3130 | 2.3159 |

Table 4. The DEF for 10 mm depth target location

| Energy, kV | 60 | | 120 | | 180 | | 250 | |
|---|---|---|---|---|---|---|---|---|
| Conc., mM | PClab | Geant4 | PClab | Geant4 | PClab | Geant4 | PClab | Geant4 |
| 0,003 | 1.0000 | 1.0000 | 1.0000 | 1.0000 | 1.0000 | 1.0000 | 1.0000 | 1.0000 |
| 0,006 | 1.0000 | 1.0000 | 1.0000 | 1.0000 | 1.0000 | 1.0000 | 1.0000 | 1.0000 |
| 0,009 | 1.0000 | 1.0000 | 1.0000 | 1.0000 | 1.0000 | 1.0000 | 1.0000 | 1.0000 |
| 0,012 | 1.0000 | 1.0000 | 1.0000 | 1.0000 | 1.0000 | 1.0000 | 1.0000 | 1.0000 |
| 0,3 | 1.0000 | 1.0000 | 1.0044 | 1.0066 | 1.0081 | 1.0183 | 1.0162 | 1.0138 |
| 0,6 | 1.0000 | 1.0000 | 1.0115 | 1.0125 | 1.0119 | 1.0123 | 1.0150 | 1.0096 |
| 0,9 | 1.0000 | 1.0273 | 1.0187 | 1.0194 | 1.0115 | 1.0283 | 1.0108 | 1.0201 |
| 1,2 | 1.0000 | 1.0366 | 1.0198 | 1.0284 | 1.0322 | 1.0398 | 1.0219 | 1.0231 |
| 3 | 1.0577 | 1.0482 | 1.0713 | 1.0723 | 1.0600 | 1.0655 | 1.0246 | 1.0309 |
| 6 | 1.1440 | 1.1303 | 1.1459 | 1.1477 | 1.1197 | 1.1235 | 1.0552 | 1.0544 |
| 9 | 1.2292 | 1.2137 | 1.2138 | 1.2185 | 1.1738 | 1.1713 | 1.0947 | 1.0970 |
| 12 | 1.2995 | 1.2973 | 1.2806 | 1.2997 | 1.2212 | 1.2396 | 1.1428 | 1.1419 |
| 30 | 1.8057 | 1.7924 | 1.724 | 1.7293 | 1.5692 | 1.5618 | 1.3570 | 1.3554 |
| 60 | 2.5975 | 2,5899 | 2.4218 | 2.4515 | 2.1234 | 2.1228 | 1.6972 | 1.6990 |
| 90 | 3.3736 | 3.3506 | 3.1367 | 3.1643 | 2.6708 | 2.6697 | 2.0651 | 2.0653 |
| 120 | 4.2406 | 4.4234 | 3.8270 | 3.8598 | 3.2061 | 3.2456 | 2.3870 | 2.3885 |

Table 5. The DEF for 50 mm depth target location

| Energy, kV | 60 | | 120 | | 180 | | 250 | |
|---|---|---|---|---|---|---|---|---|
| Conc., mM | PClab | Geant4 | PClab | Geant4 | PClab | Geant4 | PClab | Geant4 |
| 0,003 | 1.0000 | 1.0000 | 1.0000 | 1.0000 | 1.0000 | 1.0000 | 1.0000 | 1.0000 |
| 0,006 | 1.0000 | 1.0000 | 1.0000 | 1.0000 | 1.0000 | 1.0000 | 1.0000 | 1.0000 |
| 0,009 | 1.0000 | 1.0000 | 1.0000 | 1.0000 | 1.0000 | 1.0000 | 1.0000 | 1.0000 |
| 0,012 | 1.0000 | 1.0000 | 1.0000 | 1.0000 | 1.0000 | 1.0000 | 1.0000 | 1.0000 |
| 0,3 | 1.051 | 1.0064 | 1.0161 | 1.0265 | 1.0209 | 1.0283 | 1.0005 | 1.0098 |
| 0,6 | 1.0133 | 1.0175 | 1.0236 | 1.03178 | 1.0145 | 1.0265 | 1.0093 | 1.0127 |
| 0,9 | 1.0154 | 1.0314 | 1.0355 | 1.0495 | 1.0231 | 1.0214 | 1.0086 | 1.0083 |
| 1,2 | 1.0232 | 1.0387 | 1.0295 | 1.0551 | 1.0226 | 1.0386 | 1.0110 | 1.0122 |
| 3 | 1.0573 | 1.0500 | 1.0839 | 1.0952 | 1.0643 | 1.0702 | 1.0369 | 1.0355 |
| 6 | 1.1400 | 1.1356 | 1.1577 | 1.1646 | 1.1196 | 1.1161 | 1.0859 | 1.0724 |
| 9 | 1.2250 | 1.2203 | 1.2224 | 1.2460 | 1.1769 | 1.1836 | 1.1066 | 1.1136 |
| 12 | 1.2959 | 1.2995 | 1.2890 | 1.3054 | 1.2258 | 1.2390 | 1.1510 | 1.1537 |
| 30 | 1.7933 | 1.7863 | 1.7206 | 1.7360 | 1.5853 | 1.58298 | 1.3432 | 1.3452 |
| 60 | 2.5613 | 2.5647 | 2.4204 | 2.4872 | 2.1475 | 2.1340 | 1.6996 | 1.6911 |
| 90 | 3.3039 | 3.3075 | 3.2333 | 3.2134 | 2.7374 | 2.7056 | 2.0532 | 2.0518 |
| 120 | 4.0713 | 4.0002 | 3.946 | 3.9350 | 3.2171 | 3.2306 | 2.3615 | 2.3664 |

The graph 19 shows a linear increase of DEF at increasing concentration. Based on the orthovoltage X-rays energy considered, the highest degree of dose enhancement occurs for 60 kV and higher DEA concentration.

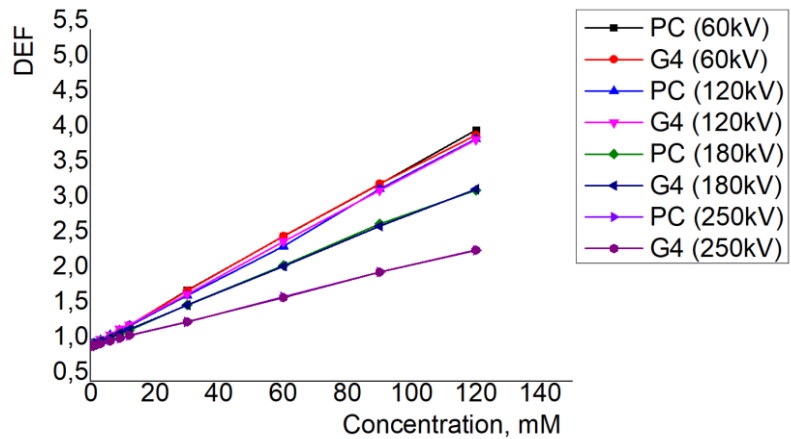

Figure 19. The dependence of DEF on cisplatin concentration at X-ray energies of 60 kV to 250 kV.

Figure 20a presents a comparison between PDD from photon beams with energies of 6 MV, 1.25 MV(Co-60), 60 kV and 60 kV with cisplatin is shown. Also, figure 20b shows a comparison of the PDD between the 6MV, 1.25 MV(Co-60), 250 kV and 250 kV with cisplatin.

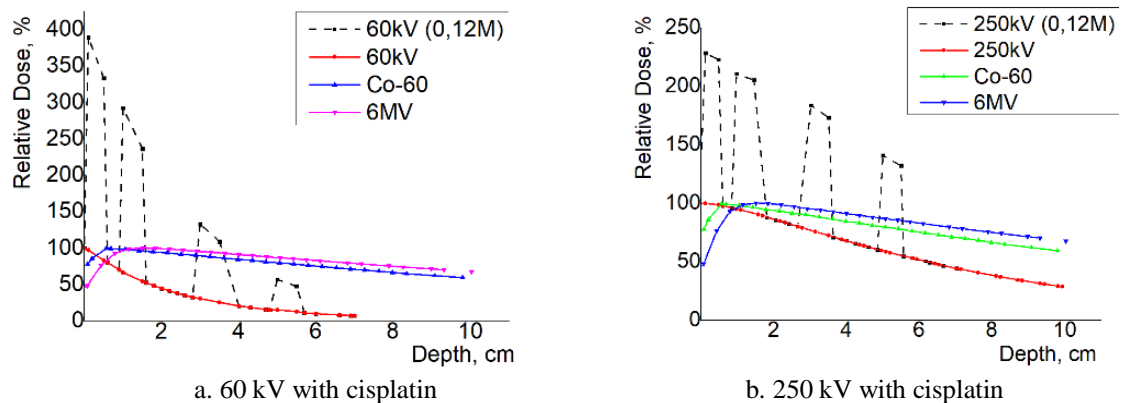

a. 60 kV with cisplatin  b. 250 kV with cisplatin

Figure 20. Comparison between PDD from different energy photon beams

According to the graph, that in the field of DEA accumulation there is a sharp increase in dose not only for surface targets, but also for deeply targets located. For example, in the case of the cisplatin presence in the target and irradiation by a photon beams with an energy of 250 kV a dose accumulation is higher than when irradiated with photon beams with an energy of MV and 1.25 MV(Co-60), and even at a depth target location of 5 cm is observed.

## 4 Conclusion

This research considers potential dose increase in target due to cisplatin (Pt) concentration and radiation type. In general, Monte-Carlo results showed that dose escalation in target at orthovoltaic x-ray and megavolt photon beams can be due photon-capture therapy. For X-rays photon beams the DEF was higher than for megavoltage photon beam generated from linear acceleration. For 60 kV x-rays, a tremendous DEF was seen, ranging of almost 4.3 at the highest Pt concentration. The dose enhancement is somewhat less for with increasing energy up to 250 kV. These differences become important with very high DEA concentrations to 120 mM in tumor.

We should also notice, that the DEF is higher in linacs without flattening filter than in linacs with flattening filter. For the 6 and 10 MV photon beams with flattening filter the DEF to 1.077 and 1.031 was seen. As well as 6 and 10 MV without flattening filter - DEF equal to 1.086 and 1.089 was observed. It can be assumed that this phenomenon is due to a cause that the flattening filter system absorbs low-energy photon beam. The same time these low-energy photons cause a photoelectric effect in FFF systems and corresponding dose increase in the target is observed.

Perhaps from the point of view of clinical significance the DEF values obtained for the 6 and 10 MV are not important. However, from the point of view of considering the error of dose adjustment these factors can be taken into account, for example, with simultaneous chemoradiation treatment or the other DEA introduction.

This way, Monte-Carlo simulation showed that the dose-enhancing factor increases for higher values of cisplatin concentrations and lower photon energy.

**References:**

[1] R. Baskar, K.A. Lee, R.Yeo, et al., *Cancer and Radiation Therapy: Current Advances and Future Directions*, Int. J. Med. Sci. **9** (2018) 193.

[2] I.N. Sheino, P.V. Izhevskij, A.A. Lipengolts, et al., *Development of binary technologies of radiotherapy of malignant neoplasms: condition and problems*, Bulletin of Siberian Medicine. **16** (2017) 192.

[3] V. F. Hohlov. V. N. Kulakov. I. N. Sheino. et al., *Method of photon-capture therapy of tumors*, Burnazyan Federal Medical Biophysical Center. RU Patent No. 2270045 (2006).

[4] Jain S. Coulter JA. Hounsell AR. et al., *Cell-specific radiosensitization by gold nanoparticles at megavoltage radiation energies,* Int J Radiat Oncol Biol Phys. **79** (2010) 531.

[5] D. Chithrani, S. Jelveh, F. Jalali, et al., *Gold nanoparticles as radiation sensitizers in cancer therapy*, Radiat Res. **173** (2010) 719.

[6] R. Berbeco. H. Korideck. W. Ngwa, et al., *TU-C-BRB-11: In Vitro Dose Enhancement from Gold Nanoparticles under Different Clinical MV Photon Beam Configurations*, AAPM. **39** (2012) 3900.

[7] Gorodetsky R., Levy-Agababa F., Mou X., et al., *Combination of cisplatin and radiation in cell culture: effect of duration of exposure to drug and timing of irradiation*, Int. J. Cancer. **75** (1998) 635.

[8] Chougule P.B., Suk S., Chu Q.D., *Cisplatin as a radiation sensitizer in the treatment of advanced head and neck cancers. Results of a phase II study*, Cancer. **74** (1994) 1927.

[9] K. Skov., *Interaction of Platinum Drugs With Clinically Relevant X-Ray Doses in Mammalian Cells: A Comparison of Cisplatin, Carboplatin, Iproplatin, and Tetraplatin,* Int. J. Radiat. Oncol. Biol. Phys. **20** (1991) 221.

[10] GEANT4 collaboration, S. Agostinelli et al., *GEANT4: A Simulation toolkit*, Nucl. Instrum. Meth. **506** (2003) 250.

[11] Bespalov V.I., *Computer lab (Version 9.9). Program description, program manual*, Tomsk Polytechnic University, Russia, (2016), pg. 126.

[12] Yahya Abadi A., Ghorbani M., Mowlavi A. A., et al., *A Monte Carlo evaluation of dose enhancement by cisplatin and titanocene dichloride chemotherapy drugs in brachytherapy with photon emitting sources*, Australas Phys Eng Sci Med. **37** (2014) 327.

[13] Bakhshabadi, M., Ghorbani, M., & Meigooni, A. S., *Photon activation therapy: a Monte Carlo study on dose enhancement by various sources and activation media.* Australas Phys Eng Sci Med. **36** (2013) 301.

[14] Roeske, J. C., Nuñez, L., Hoggarth, M., et al., *Characterization of the Theoretical Radiation Dose Enhancement from Nanoparticles*. Technol Cancer Res Treat. **6** (2007) 395.

[15] Hao, Y., Altundal, Y., Moreau, M., et al*., Potential for enhancing external beam radiotherapy for lung cancer using high-Z nanoparticles administered via inhalation.* Phys Med Biol. **60** (2015) 7035.

[16] Cho S, Jeong J, Hyeong K*, Monte Carlo simulation study on dose enhancement by gold nanoparticles in brachytherapy*. J Korean Phys Soc. **56** (2010) 1754.